# Coincident Measurement of the Energy Spectra of Doppler-Shifted Annihilation Gamma Quanta and Positron-Induced Secondary Electrons

R W Gladen[1, a)], V A Chirayath[1], A J Fairchild[1], N K Byrnes[1], A R Koymen[1], and A H Weiss[1]

[1]*Department of Physics, University of Texas at Arlington, Arlington, Texas 76019, USA*

[a)] Corresponding author: randall.gladen@uta.edu

**Abstract.** Preliminary results are presented from a new positron beam system currently under development at the University of Texas at Arlington for the coincident energy measurement of Doppler-shifted annihilation quanta and positron-induced Auger electrons. We report data based on an analysis of the pulses resulting from the detection of positron induced secondary electrons by a micro-channel plate detector in coincidence with the pulses resulting from the detection of associated annihilation gamma rays in a NaI(Tl) gamma detector.

## INTRODUCTION

We present data obtained in the process of developing a new apparatus designed for the coincident measurement of the spectra of Doppler broadened annihilation gamma rays [1,2] and the time-of-flight of annihilation-induced Auger electrons [2,3] When fully developed, the apparatus will make use of a very low energy beam to measure the time of flight of annihilation-induced Auger electrons in coincidence with the energy of Doppler broadened annihilation gamma rays, as measured using a high purity germanium detector. When used to detect Auger electrons, the system will also be configured so as to prevent positron impact-induced secondary electrons from reaching the electron detector. However, in the measurements reported here, the system was deliberated configured so as to generate and allow the transmission of a large number of secondary electrons and a correspondingly large number of electron detector pulses for the purpose of calibrating the detection and digital signal processing systems of the new apparatus. The data shown in this paper were obtained using a NaI(Tl) detector due to the lack of an available HPGe detector at the time the data was collected. While the data were obtained primarily for the objective of deriving signals for use in the development of software for the analysis of the digitized signals obtained from the gamma and electron detectors, the information obtained from the analysis provided insights into how the new system could be used in measurements of the energy spectra of the annihilation gamma rays from triplet positronium (Ps).

## EXPERIMENTAL

The positrons are generated by a Na-22 source, moderated by a solid neon moderator (RGM), filtered by a saddle coil energy filter, and guided through the beamline by a series of Helmholtz coils. A set of ExB plates drifts the positrons beneath a microchannel plate (MCP) detector and then back along the axis of the beam; these plates are also responsible for drifting the resultant Auger or secondary electrons from the sample into the MCP. Following the ExB plates is a bias-able stainless steel tube (ToF tube) that provides a field-free region for the majority of the flight path of the electrons, and also allows for the deceleration or filtering of the electrons. The final chamber hosts the sample manipulator and sample, which may be biased from ground up to -25 kV. Biasing the sample to a negative voltage increases the mean energy of the positron beam, and the maximum and minimum energy of the emitted secondary electrons (Fig. 1).

Two different electronic configurations were used. An analog time-of-flight spectra of annihilation induced Auger electrons was obtained using a configuration similar to that used by Mukherjee et al. [4] with the sample biased at -3V (resulting in an incident positron energy of ~8 eV) and the time of flight tube biased at -15 V (see Fig. 2) to provide preliminary data for the purpose of testing the electron and gamma detection systems. The analog electronics consisted of two fast pre-amplifiers for the NaI(Tl) gamma detector and the MCP electron detector, followed by constant fraction discriminators (CFD) for both, and a time delay for the NaI(Tl) pulse. This allowed the collection of the lower count rate MCP pulses as the starting pulse in the time-to-amplitude converter to reduce total dead time. Finally, the time-of-flight spectrum was constructed using a multi-channel analyzer (MCA).

Two dimensional spectra of annihilation gamma ray energies measured in coincidence with the time interval between the detection of the annihilation gammas and the detection of secondary electrons emitted from the sample were obtained using a digital data collection system [5] based on a Lecroy model HD 4096 oscilloscope. In these the sample was biased at -1000 V (resulting positrons hitting the sample at ~1003eV) and the TOF tube was at ground (resulting in lowest and highest energies of secondary electrons as the traversed the TOF tube of ~1003 eV to ~2006 eV respectively). The signal from the MCP was first amplified using a Mini-Circuits fast preamplifier and then fed into one input of the digital scope. The signal from the anode output of the NaI(Tl) detector was fed directly into the second input of the digital scope without amplification.

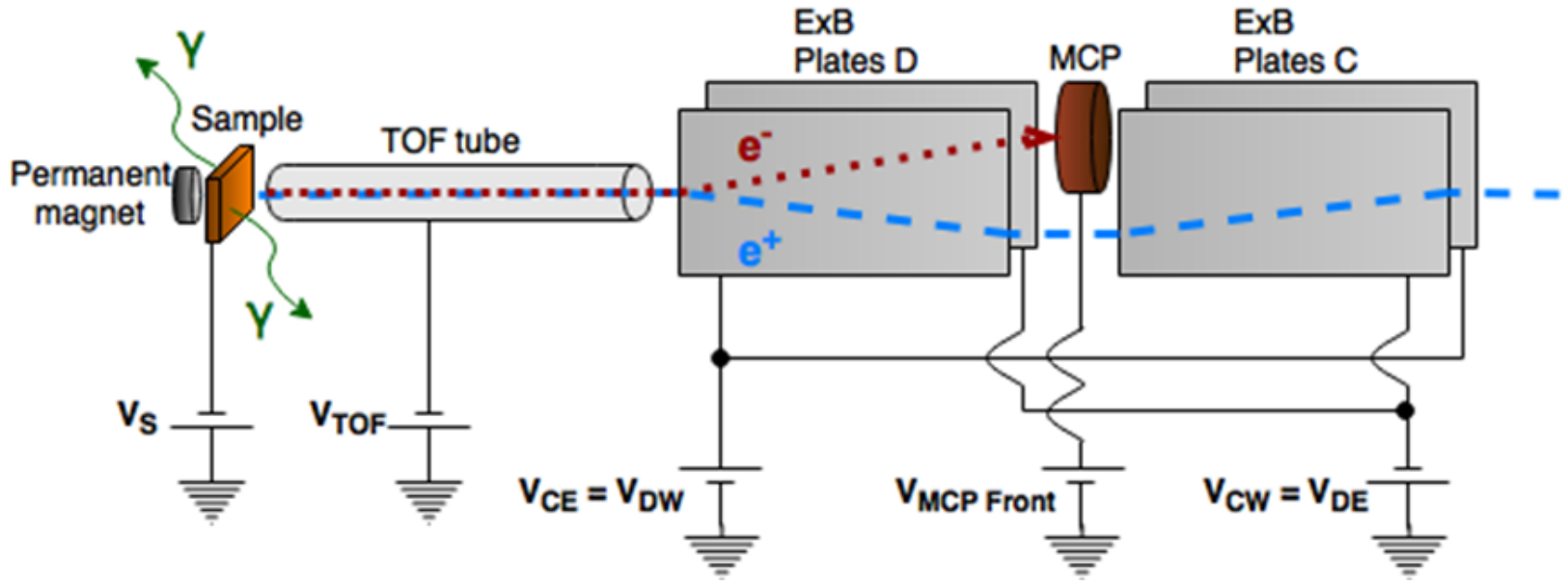

(a)

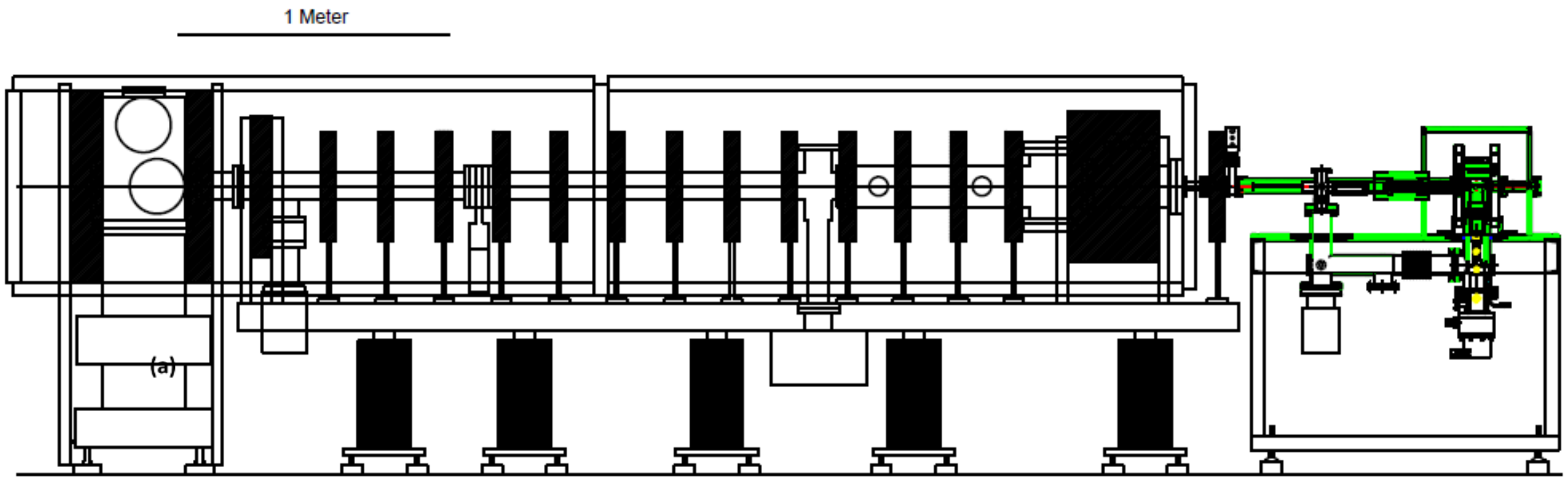

(b)

**FIGURE 1.** (a) Internal diagram exhibiting the ion optics of the beamline, neglecting the rare gas moderator. (b) External drawing of the apparatus, with the rare gas moderator on the right end of the drawing.

# ANALYSIS

The first Auger electron time-of-flight spectrum of the new beamline was collected by analog nuclear electronics (Fig. 2). However, the full analysis of the detector pulses in coincidence required the development of software capable of providing real-time construction of the spectra.

## Analog Spectrum

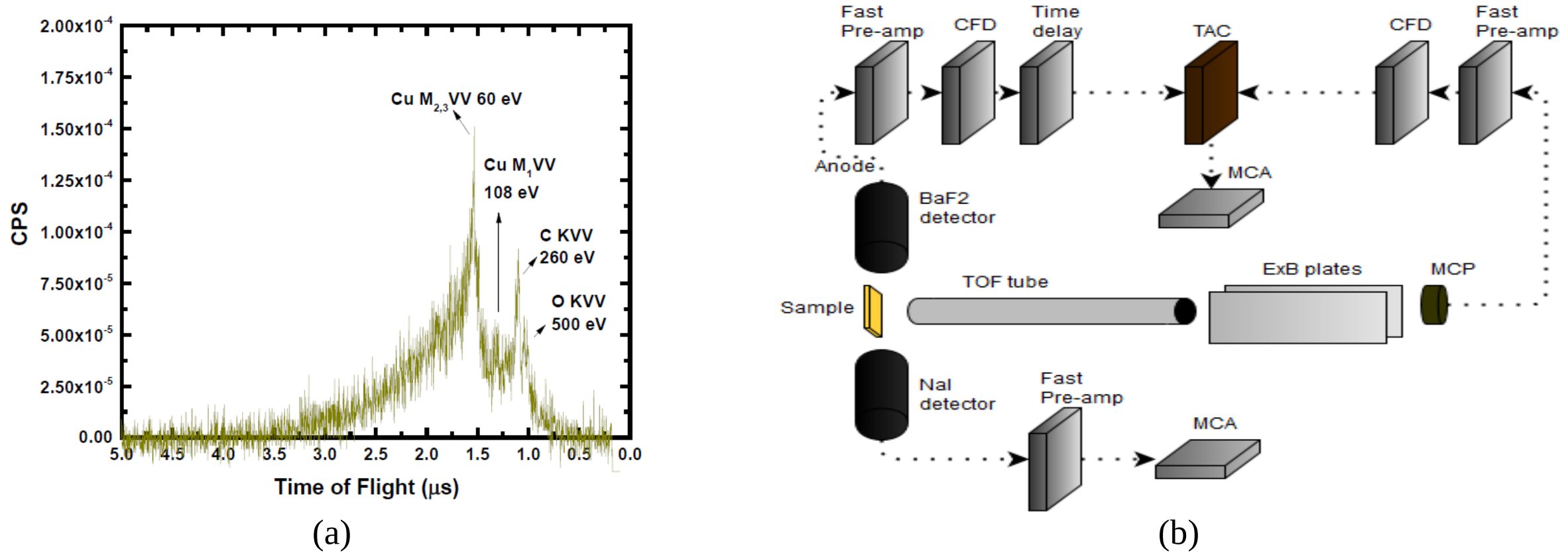

**FIGURE 2.** (a) A time-of-flight spectrum of positron-annihilation induced Auger electrons from a polycrystalline copper surface which had been sputter-cleaned and subsequently exposed to residual gas. Sample bias: -3 V. ToF tube bias: -15 V. Beam energy: ~8 eV. (b) A diagram of the timing electronics used to perform PAES. It is possible to completely remove the secondary background by applying a bias to the ToF tube.

## Software

The software used in the analysis of the detector pulses was written in MATLAB and allows the re-analysis of a single experiment by recording all of the coincident pulses, rather than just the particular features of pulses as might be done with a microcontroller or FPGA. This was required in order to easily adapt the software to the continual development of the beamline and the techniques used.

As the pulses from the gamma detector and MCP detector are collected (or after they have already been collected), they are sent to an FFT filter function in order to remove noise inherent to the lab environment. Following this, each voltage pulse in the MCP waveform is counted and sent to a digital CFD and the pick-off time is calculated. The corresponding gamma pulse is sent through a rise-time filter and an extrapolated leading edge timing (ELET) function for the calculation of the pick-off time. This same pulse is also sent to a shaping function [6] and a spectrum stabilizer function for the calculation of the photon energy. Finally, the time difference between the two pulses and the photon energy is used to construct a two-dimensional coincidence matrix.

# PRELIMINARY RESULTS

Initial results of the new apparatus and software include the collection of positron impact-induced secondary electrons in coincidence with the annihilation gamma resulting from the eventual (< 1 ns) annihilation of the positron with an electron in the material. Due to the time-of-flight measurement being dependent on the time of arrival of the gamma photon, the secondary spectrum is distorted by the lifetime of ortho-positronium (triplet Ps). This creates a "positronium tail" that is characteristic of the lifetime of triplet Ps near the surface of the material (Fig. 3b).

By integrating along either axis of Fig. 3a, the individual time-of-flight and gamma spectra are constructed (Fig. 3b-c). In addition, it is possible to select a region of either spectra and use that region as a window from which features due to particular events may be extracted from the corresponding spectrum. For example, selecting the high momentum side of the 511 keV peak in the gamma spectrum as a window results in a timing spectrum that is representative of the time-of-flight spectrum of positron impact induced secondary electrons and which is largely free of events associated with the delayed gamma signal resulting from the annihilation of long-lived triplet Ps. In contrast, selecting the valley between the photo-peak and the Compton edge of the gamma spectrum as a window will result in a timing spectrum with an enhancement in the number of events associated with the delayed annihilation signal associated with triplet Ps (Fig. 3d).

This concept can also be applied in reverse: selecting the tail of the timing spectra associated with the delayed annihilation signal from triplet Ps as a window results in a gamma spectrum with enhanced 3-gamma contributions, with a valley-to-peak ratio of 46%, and selecting events in the timing spectrum corresponding to secondary electrons from the peak of the secondary electron distribution that are detected in coincidence with the prompt annihilation gammas from the sample results in a gamma spectrum with slightly reduced 3-gamma contributions, with a valley-to-peak ratio of 28% (Fig. 3e).

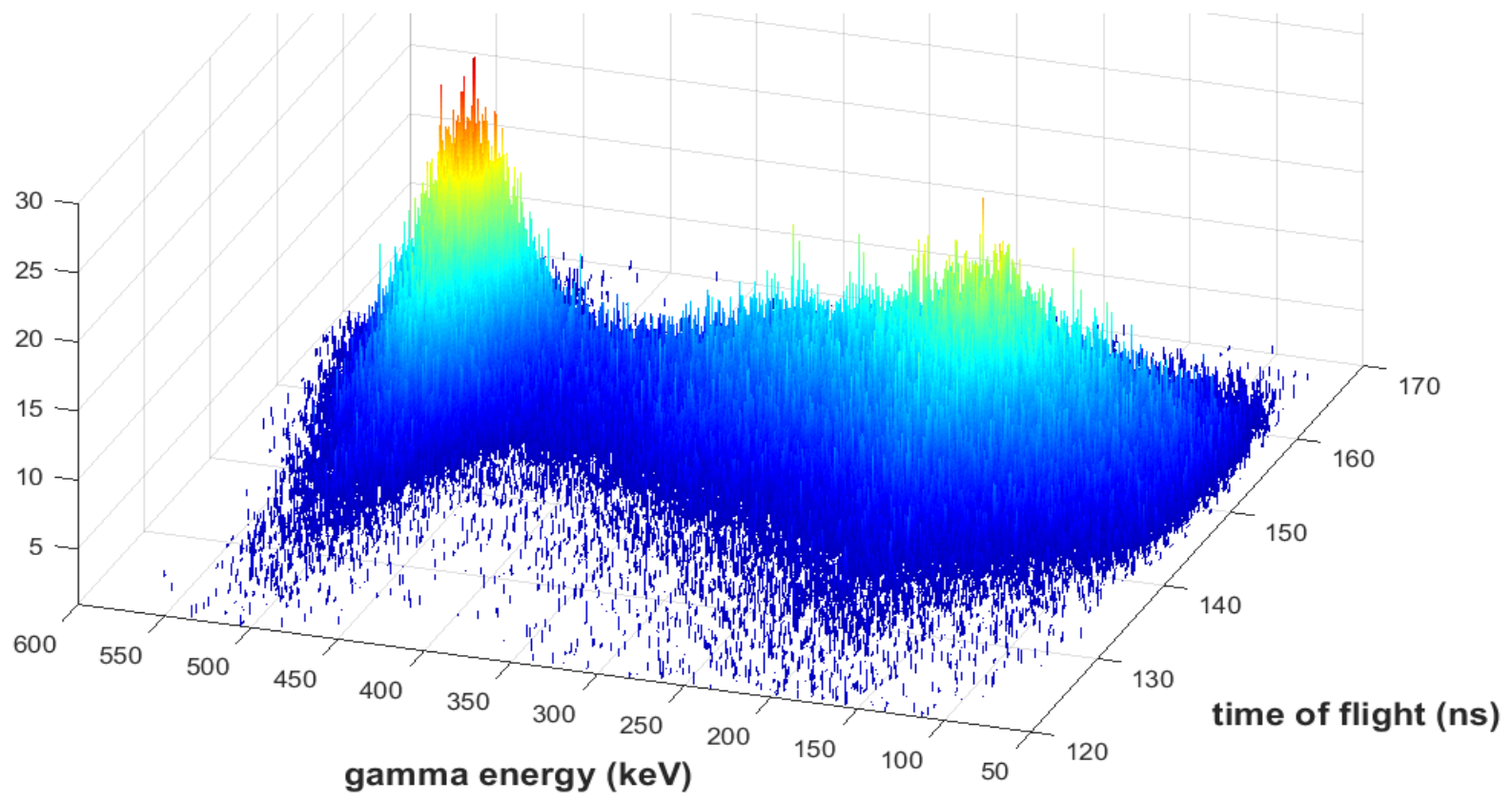

(a)

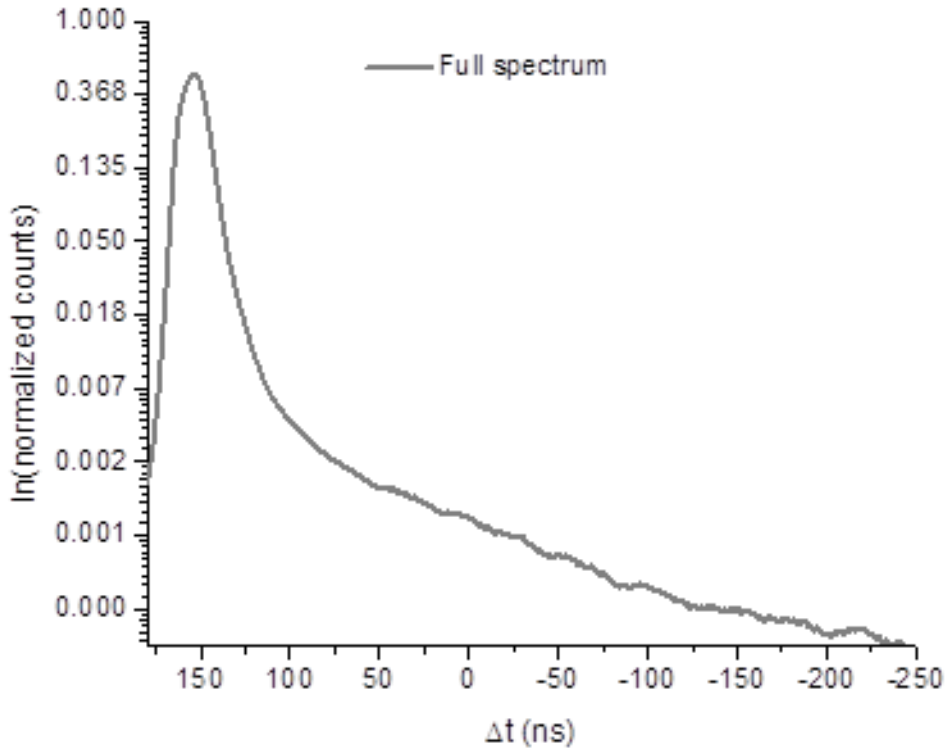

(b)

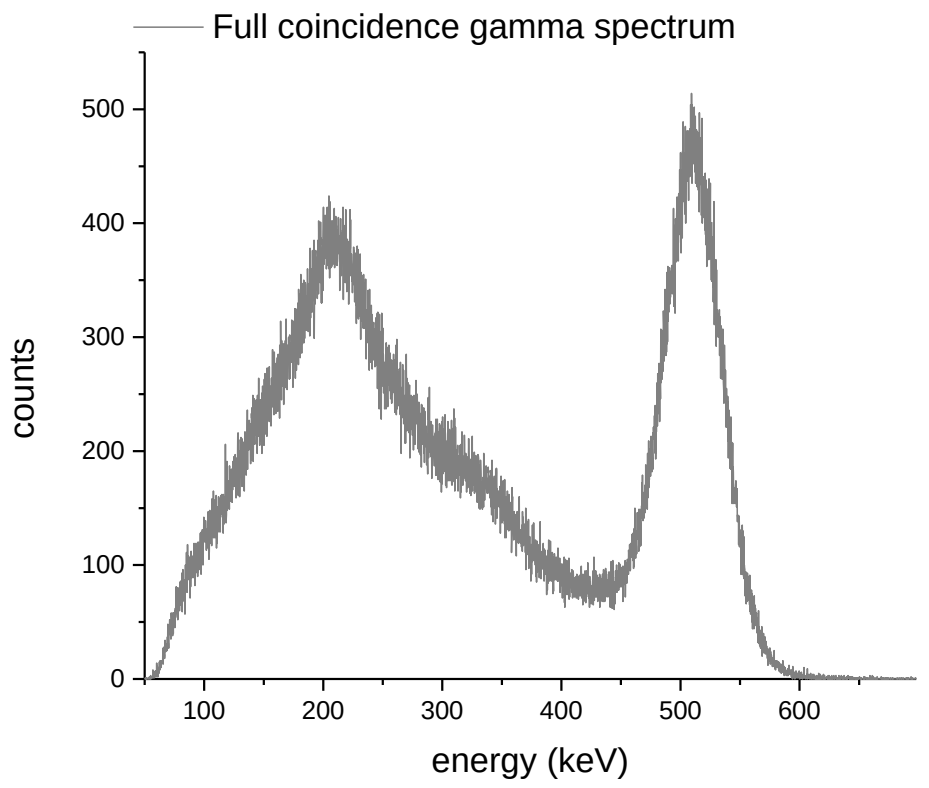

(c)

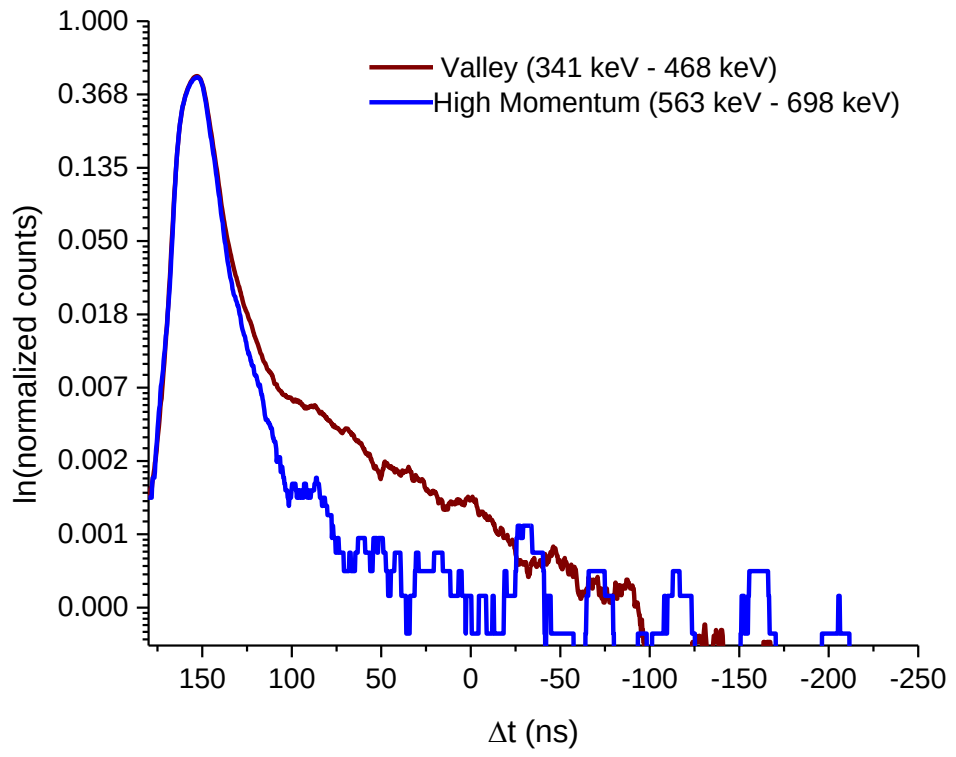

(d)

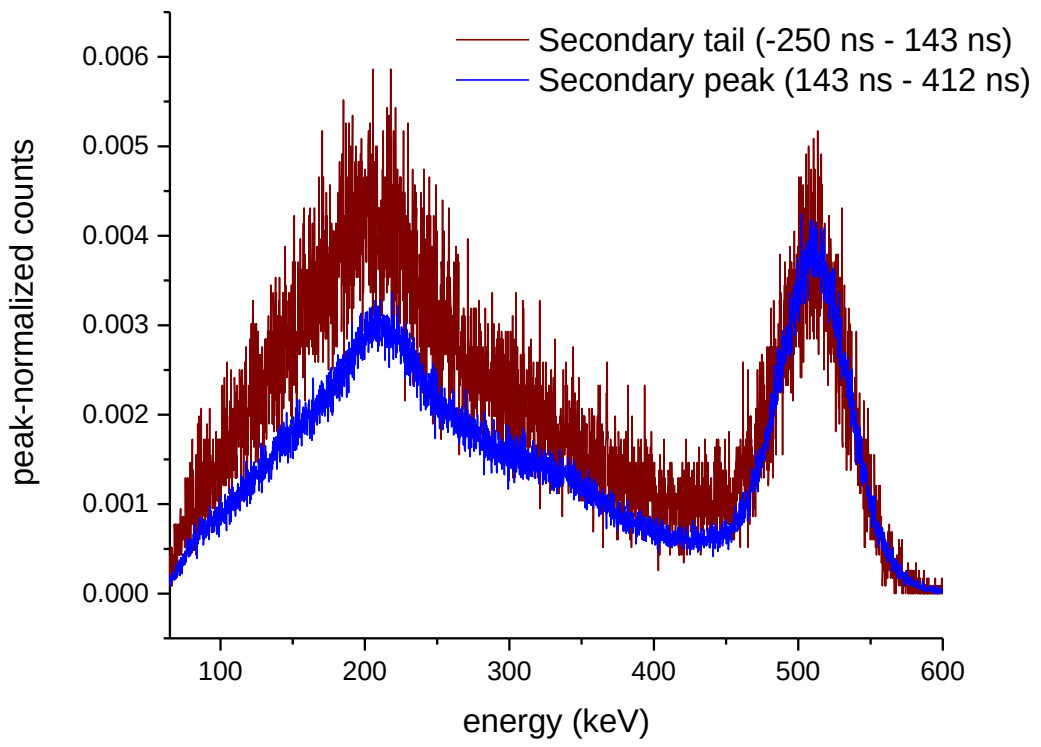

(e)

**FIGURE 3.** (a) A 2-dimensional histogram of gamma-electron coincidence from bi-layer graphene on a substrate of polycrystalline copper. Gamma photons were collected by the same NaI(Tl) detector used for the electron time of flight spectrum. Sample bias: -1 kV. Moderator bias: +1 V. Initial beam energy: ~2.5 eV. (b) The integrated electron-gamma timing spectrum. (c) The integrated gamma energy spectrum. (d) Secondary spectra in coincidence with particular energy ranges of gamma photons. (e) Gamma spectra in coincidence with electrons within a particular time of flight range.

## SUMMARY

We've reported a technique for the coincident measurement of the energy spectra of Doppler-broadened annihilation gammas and positron-induced secondary electrons. With this technique, secondary electron time-of-flight spectra free of the positronium tail or with enhanced positronium contributions may be constructed. Further, gamma spectra free of triplet Ps contributions or with enhanced triplet Ps contributions may also be constructed. The ability to construct electron time-of-flight and gamma spectra in coincidence with each other allows future experiments with the capability to correspond annihilation events with the resulting electron emission processes. Planned experiments include the collection of a gamma-Auger coincidence spectra from the surface of materials known for positron stimulated ion desorption [7]. In addition, by analyzing the Doppler-broadened gamma spectrum in coincidence with a range of electrons in the Auger spectrum that may correspond to a multi-Auger process, it may be possible to determine the particular annihilation event that results in multi-Auger emission [8]. Finally, we plan further experiments regarding the VVV Auger process in which the VVV region in the Auger electron spectrum will be measured in coincidence with the region in the gamma spectrum corresponding to valence annihilations [9,10].

## ACKNOWLEDGMENTS

This work was supported by NSF grants DMR 1508719 and DMR 1338130, and Welch Foundation grant No. Y-1968-20180324. The authors would like to thank Professor Yasuyuki Nagashima for collaboration and discussion, from which planned experiments involving the aforementioned technique were derived.